\documentstyle[aps,psfrag,graphicx,preprint,tighten,eqsecnum,floats,epsf,epsfig,rotate,prd,here]{revtex}
\newcommand{\be}{\begin{eqnarray}}
\newcommand{\ee}{\end{eqnarray}}
\newcommand{\sbe}{\begin{eqnarray*}}
\newcommand{\see}{\end{eqnarray*}}
\newcommand{\pslash}{\not\!p}

\begin{document}
\draft
\vskip 4mm\preprint{
Preprint No. ADP -00-54/T434}


\title{Regularization-independent study of renormalized non-perturbative quenched QED}

\author{
        Ay{\c s}e K{\i}z{\i}lers{\" u}
            \footnote{E-mail:~akiziler@physics.adelaide.edu.au},
        Tom Sizer
            \footnote{E-mail:~tsizer@physics.adelaide.edu.au}
          and
        Anthony G.\ Williams
            \footnote{E-mail:~awilliam@physics.adelaide.edu.au}
        \vspace*{5mm}
        }

\address{
   Special Research Centre for the Subatomic Structure of Matter and \\
   Department of Physics and Mathematical Physics, \\
   Adelaide University, 5005, Australia
   \vspace*{2mm}
        }

%
%

\maketitle

\begin{abstract}
  A recently proposed regularization-independent method is used for the
  first time to 
  solve the renormalized fermion Schwinger-Dyson equation numerically 
  in quenched QED$_4$. The Curtis-Pennington vertex is used to illustrate 
  the technique and to facilitate comparison with previous calculations 
  which used the
  alternative regularization schemes of modified ultraviolet cut-off and dimensional 
  regularization. Our new results are in excellent numerical agreement with these, 
  and so we can now conclude with confidence that there is no residual regularization 
  dependence in these results. 
  Moreover, from a computational point of view the regularization
  independent method has enormous advantages, since all integrals are absolutely 
  convergent by construction, and so do not mix small and arbitrarily large momentum scales. 
  We analytically predict power law behaviour 
  in the asymptotic region, which is confirmed numerically with high precision. 
  The successful demonstration of this efficient new technique opens the way for
  studies of unquenched QED to be undertaken in the near future. 
\end{abstract}
\newpage
\section{INTRODUCTION}
\label{sec_intro}
The divergences inherent in quantum field theories have plagued
physicists for years. 
The infinities are removed in a two-step process; first the
divergences are controlled by a regulator, then the regulator is removed
using the renormalization procedure to obtain finite,  renormalization-independent 
physical quantities. Depending on the type of regulator introduced, different
problems can occur. It is useful to discuss the difficulties that
arise in the Schwinger-Dyson equations (SDE's)~\cite{RW_Review}, since 
these are identical to the difficulties encountered in perturbation theory and 
because 
we will be using them as a tool to study
non-perturbative QED.  In the literature SDE's in
quenched QED$_4$ are commonly studied by using the ultraviolet cut-off
regularization scheme~\cite{qed4_hw_etal0}.  
One difficulty faced is 
that the use of an ultra-violet (UV) cut-off $\Lambda$ to regulate the 
integration will in general lead to an explicit violation of gauge 
covariance~\cite{DMR}.  Because
this regularization scheme does not respect translation invariance in the
loop-momentum integration, it
will lead to an explicit gauge-covariant violating contribution in the
result, {\em even after} $\Lambda$ is taken to infinity.  More
precisely, this violation of gauge invariance has been observed in quenched
QED calculations employing the Curtis-Pennington (CP)~\cite{mike2} electron-photon
vertex and may be traced back to a certain, logarithmically divergent,
4-dimensional momentum integral which vanishes because of rotational
symmetry at all $\Lambda < \infty$, but leads to a finite contribution
for $\Lambda \rightarrow \infty$.  It is this discontinuous behaviour
as a function of $\Lambda$ which complicates correct numerical
renormalization with this regulator.  Incorrect results will be
obtained unless care is taken to identify and remove `gauge-covariance violating
terms'~\cite{qed4_hw_etal1}-~\cite{qed4_hw_etal2}.  In its favour,
cut-off regularization is computationally economical compared to dimensional 
regularization, and gives accurate answers after gauge-covariant violating 
terms are removed.

On the other hand, SDE studies implemented using a gauge-invariant
regularization scheme, such as dimensional 
regularization~\cite{qed_dim_reg}, 
will not have such a problem. However it will cause the generation of dynamical
mass for all coupling constants in $D \neq 4$, instead of only above
critical coupling in $D=4$.  In non-perturbative studies this scheme
is computationally more demanding since a careful  and time-consuming
removal of the regulator must be performed involving an extrapolation of
many high-precision solutions for different $\epsilon$ to $\epsilon = 0$.
Additionally the accuracy of the results is, in practice, more limited
because the integration range neccessarily  extends to infinity.

In this paper we will be employing for the first time the regulator-independent method 
recently proposed by K{\i}z{\i}lers{\" u} et.al.~\cite{KSSW}. This method deals
with the renormalized quantities only, 
as the regulator is removed analytically. The dependence on the mass scale
introduced by the regulator is traded for the momentum scale $\mu$ at
which the theory is renormalized {\it before} performing any numerical
calculations. In this way dimensional regularization (or any other
regulator which doesn't violate gauge covariance) can be used and the regulator 
removed analytically {\it before} any numerical calculations are begun.
More importantly from a numerical pointof view, removal of the
regulator means that no longer does one have to solve integral
equations involving mass scales of vastly different orders of
magnitude (and then, in addition, take a limit in which one of the
scales goes to infinity). Rather, the important scales in the problem
become scales of {\it physical} importance, such as the renormalized
mass $m_\mu$ and the renormalization scale $\mu$ at which this mass is
defined. It is therefore to be expected that the dominant
contributions to any integrands will be from a finite region
of momenta. This feature is generic; i.e., it is independent of the
particular vertex ansatz that one makes use of, and it remains a valid
consideration for an arbitrary renormalizable field theory.  

The
regularization-independent approach is computationally very economical 
and very accurate; however the price  paid is that one loses all contact 
with the bare theory. In particular, because of this, 
one cannot study dynamical chiral symmetry breaking in this approach 
by simply setting the bare mass $m_0=0$ and investigating at what 
value of coupling the dynamical fermion mass is generated. 
It is in this crucial point we differ from the analysis in 
Ref.~\cite{CP92} where, within quenched
QED, a removal of the regulator was attempted in a manner which has
some similarity to what we do here. Indeed, as emphasized 
by Miransky (Refs.~\cite{noreg}, \cite{FGMS_Review} as well as 
chapter 10.7 of Ref.~\cite{Miransky_book}),
some care needs to be taken with the treatment of the bare mass while
removing the regulator in order to avoid drawing incorrect conclusions
about the presence or absence of dynamical chiral symmetry breaking in
gauge theories. In the present approach this problem is avoided.
Dynamical chiral symmetry breaking, both explicit as well as dynamical, is
characterized by a non-zero renormalized mass $m_\mu$, and hence
this quantity in itself cannot distinguish between these two
possibilities. In section~\ref{sec:asymp}, we use (within quenched
QED$_4$) the appearance of oscillations
in the renormalized fermion mass function~\cite{noreg} as the indicator of
the onset of dynamical symmetry breaking. All previous studies have 
confirmed that the occurrence of dynamical mass generation and UV 
oscillations coincide in quenched QED$_4$. 
 
In this
paper we demonstrate this approach numerically for the specific case of 
quenched QED$_4$ using the CP vertex 
to test its validity against previous works.  Our goal is
to ultimately implement this technique in studies of unquenched QED$_4$, 
which have now become numerically tractable because of this approach. 

In section~\ref{sec_reg_indep}
 we formulate the regularization-independent method
for renormalized SDE for quenched QED in 4-dimensions with an
arbitrary covariant gauge. Section~\ref{sec:sec3}  discusses the large
momentum behaviour of the fermion propagator  and gives its
analytical form. We go on to solve SDE's numerically in Euclidean space for
the fermion wave-function renormalization function and the mass function
employing the CP vertex. The integration range in these equations is taken 
to infinity analytically and subsequently evaluated using a combination of 
numerical and analytic results. We do this extrapolation to infinity to 
facilitate comparison with previous results at very high momentum scales 
and to compare these results with their asymptotic forms. We demonstrate
numerically to high precision the agreement between the regularization 
independent approach and the modified cut-off and dimensional regularization 
schemes. Finally, in section~\ref{sec:conclusions}
we summarize our results and conclude.

\section{REGULARIZATION-FREE FORMALISM IN QUENCHED QED$_4$}
\label{sec_reg_indep}

\begin{figure}[H]
\begin{center}
~\epsfig{file=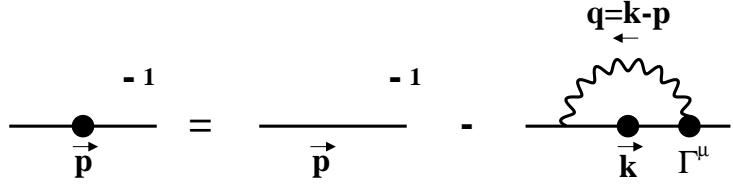,width=300pt}
\end{center}
\caption{The Schwinger-Dyson equation for the fermion propagator.}
\label{fig:vertex}
\end{figure}

The renormalized Schwinger-Dyson equation for the electron propagator 
(Fig.~\ref{fig:vertex}) can be formulated as 

\be
S^{-1}(\mu;p) &=& Z_2(\mu) {S^{0}}^{-1}(p)
\> - \> i Z_1(\mu) {e_0}^2 \int {d^d k \over (2 \pi)^d}\>
\Gamma^\alpha (\mu;p,k) \>  S(\mu;k) \gamma^\beta \> 
 D^0_{\alpha \beta}(\mu;q)\, , \nonumber \\
&\equiv& Z_2(\mu) {S^{0}}^{-1}(p)\> - \> i Z_1(\mu)\> \overline{\Sigma}(p)\quad.
\label{eq:mainsdf} 
\ee
\noindent
Here $ Z_1(\mu)$ and $ Z_2(\mu)$ are the vertex and fermion
wave-function renormalization constants respectively.  
These renormalization constants relate the regularized but unrenormalized 
(i.e. {\it bare}) and renormalized propagator and vertex by
\be {\Gamma}_\mu(k,p\,;\mu)
\, &=& \,
Z_1(\mu,\Lambda) \, \Gamma_\mu^{\rm bare}(k,p\,;\Lambda)\quad, \\
{S}(p\,;\mu) \, &=& \, Z_2^{-1}(\mu,\Lambda) \, S^{\rm bare}(p\,;\Lambda) \quad .  
\ee 
In quenched QED, there is no
renormalization of the electron charge
$(e_{0}^2=e_{\mu}^2=4\pi\alpha)$ and the appropriate photon propagator
is simply the tree-level form
\be
D^0_{\alpha\beta}(q)\>=\>-\frac{1}{q^2}\,
\left[
\left(g_{\alpha\beta}-\frac{q_{\alpha}q_{\beta}}{q^2}\right)
+\xi\frac{q_{\alpha}q_{\beta}}{q^2}
\right] \quad,
\ee
where $\xi$ is the covariant gauge parameter.  We use $S(\mu;p)$ 
to denote the full fermion propagator renormalized at the 
momentum scale $\mu$. It can be
expressed in terms of two scalar functions, $Z(\mu^2;p^2)$, the
fermion wave-function renormalization function, and $M(p^2)$, the mass function, by
\be S(\mu;p) \>=\>
\frac{Z(\mu^2;p^2)}{\not\!p - M(p^2)}\quad. 
\ee
Note that $S^{0}(p)\equiv 1/(\not\!p-m_0)$ is the tree-level fermion propagator 
(i.e., the bare fermion propagator in the absence of interactions).      
The full (proper) renormalized fermion-photon vertex is $\Gamma_{\mu}(\mu;k,p)$.
Multiplying the fermion SDE Eq.~(\ref{eq:mainsdf}) by $\not\!p$ and
$I$ and taking their respective spinor traces, we can separate the fermion
self-energy $\overline \Sigma(p^2)$ into Dirac odd, $\overline\Sigma_d(p^2)$,
and Dirac even, $\overline \Sigma_s(p^2)$, parts;  i.e.  $\overline \Sigma(p)=
\pslash \;\overline\Sigma_d(p^2) + \overline\Sigma_s(p^2)$. This gives
\be
Z^{-1}(\mu^2;p^2) &=&  Z_2(\mu) \>-\> Z_1(\mu) \>\overline \Sigma_d(p^2)\quad,
\label{eq: z eq}\\
M(p^2)\> Z^{-1}(\mu^2;p^2)&=&  Z_2(\mu)\> m_0\> +\> Z_1(\mu)\>\overline \Sigma_s(p^2)\;\;\;,
\label{eq: m eq}
\ee
and these self energies make use of any regularization scheme which does not violate gauge 
covariance. 
Here, the bar 
over these quantities
indicates that we have explicitly separated out the renormalization
constants $Z_i(\mu)$;  note that for notational brevity we do not indicate the
implicit dependence on $\mu^2$ of $\overline \Sigma_{d,s}(p^2)$
through the function $Z(\mu^2;p^2)$.  

Because of the Ward-Takahashi identity, which must be satisfied
by any acceptable vertex Ansatz for $\Gamma_\mu(\mu;k,p)$ and any acceptable 
renormalization scheme,  one has $ Z_1(\mu)= Z_2(\mu)$.
Making use of this fact, 
Eqs.~(\ref{eq: z eq}) and (\ref{eq: m eq}) can be rearranged as
\be
Z^{-1}_{2}(\mu) &=& Z(\mu^2;p^2)-Z(\mu^2;p^2)\>\overline \Sigma_d(p^2)
\quad ,
\label{eq: z1 eq} \\
M(p^2) \> &=& \> m_{0} \> + \> \left [M(p^2) \overline \Sigma_d(p^2)
\> + \>\overline \Sigma_s(p^2)\right]\quad.
\label{eq: m1 eq}
\ee
In order to avoid cumbersome notation, we have not explicitly
indicated functional dependence on the regulator in
Eqs.~(\ref{eq:mainsdf})--(\ref{eq: m1 eq}) but it should be understood.  
The renormalization constants 
$Z_{1,2}(\mu)$ and the
bare fermion mass $m_0$ are regulator dependent in the above equations. 
As one removes the regulator, the integrals on the right hand side of
Eq.~(\ref{eq:mainsdf}), and hence
$\overline \Sigma_d(p^2)$ and $\overline \Sigma_s(p^2)$,  
diverge logarithmically.  It is the
defining feature of a renormalizable field theory that these
divergences may be absorbed into the constants $Z_{1,2}(\mu)$ and into
the bare mass $m_0$, rendering finite, regularization-independent
 limits for $Z(\mu^2;p^2)$ and
$M(p^2)$.  However, we can make use of the $p^2$
independence of $Z_{1,2}(\mu)$ and $m_0$ in order to eliminate these 
constants from the above equations. 
The renormalization conditions for the fermion wave-function 
renormalization function and the mass function are
\be
Z(\mu^2;\mu^2) \>=\> 1 \quad\quad \mbox{and}\quad\quad
M(\mu^2) \>\equiv \> m_\mu 
\label{eq:renorm cond}\;\;\;.
\ee
Evaluating Eqs.~(\ref{eq: z1 eq}) and (\ref{eq: m1 eq})
at a second momentum which, we take to be $p^2=\mu^2$, 
and taking the difference one obtains the central equations that we 
solve here
\be
\framebox{$\begin{array}{rcl} 
\\[-3mm]
Z(\mu^2;p^2) \> &=& \> 1  \> + \> Z(\mu^2;p^2) \>\overline \Sigma_d(p^2)
\> - \> \overline \Sigma_d(\mu^2)\quad,
\label{eq:subwavefunc} \\[3mm]
M(p^2) \> &=& \> m_\mu \> + \> \left [M(p^2) \overline \Sigma_d(p^2)
\> + \>\overline \Sigma_s(p^2)\right]
\> -  \> \left[m_\mu \overline \Sigma_d(\mu^2) \> + \> \overline
  \Sigma_s(\mu^2)\right] \,\, .\\[-3mm]
\label{eq:submass}
\end{array}$}\ee
As the left hand sides of these equations
must be finite as the regulator is removed, then the right hand side must
be also, even though the individual terms on the RHS may {\it separately} diverge.  

These renormalized
equations Eq.~(\ref{eq:submass})  with the
regulator removed provide a starting point for
non-perturbative investigations which have significant advantages over
the usual treatment found in the 
literature~\cite{qed4_hw_etal0}-~\cite{qed_dim_reg}.     
In the following section we shall illustrate this
approach by turning to the example of quenched QED with
the Curtis-Pennington vertex.

\section{QUENCHED QED WITH THE CURTIS-PENNINGTON VERTEX}
\label{sec:sec3}
In this section we use the regulator-independent method
for solving the renormalized
SDE in Eq.~(\ref{eq:submass}) in an arbitrary covariant gauge. 
As usual we write 
the full vertex as a sum of the 
Ball-Chiu vertex~\cite{BallChiu} 
(longitudinal part  satisfying the Ward-Takahashi identity) and the
Curtis-Pennington term~\cite{mike2} (the transverse part 
satisfying multiplicative renormalizability of the
fermion propagator),
\be
\Gamma^\mu(\mu;k,p) &=&\Gamma^\mu_{BC}(\mu;k,p)+\tau_6(\mu;k,p)\,
\left [\gamma^{\mu}(p^2-k^2)+(p+k)^{\mu}{(\not \!k-\not\!p) }\right ]
\quad ,
\label{eq:vertex}
\ee
where $\Gamma^\mu_{BC}(\mu;k,p)$ and $\tau_6(\mu;k,p)$ are given in
appendix~\ref{ap:app1}.
After Wick rotating into Euclidean space  
and performing the angular integration, Eq.~(\ref{eq:submass}) 
can be written as
\be
Z(\mu^2;p^2)  & = &  1
-\frac{\alpha\xi}{4\,\pi}\,\int_{p^2}^{{\mu}^2}
dk^2 \frac{1}{\left[k^2+M^2(k^2)\right]}\,Z(\mu^2;k^2)\nonumber\\
&& \hspace{1.5cm}
+\frac{\alpha}{4\,\pi}\,\int_0^{\infty} 
                \frac{dk^2}{\left[k^2+M^2(k^2)\right]}
    \left [Z(\mu^2;p^2)\,I(k^2,p^2) \> - \> I(k^2,\mu^2) \right ]\, ,
\label{eq:renwavefunc}\\
M(p^2)  & = & m_\mu
+\frac{\alpha}{4\pi}\int_0^{\infty}\frac{dk^2}{\left[k^2+M^2(k^2)\right]}
\left [ J(k^2,p^2) - J(k^2,\mu^2) 
\right.\nonumber \\&&\left.\hspace{5cm}
\> + \> M(p^2)\>I(k^2,p^2) - m_\mu \>
I(k^2,\mu^2)\right ]\, ,
\label{eq:renmass}
\ee
where the kernel functions
$I(p^2,k^2)$ and $J(p^2,k^2)$ are also given in appendix~\ref{ap:app1}.
The kernels have the same form as Ref.~\cite{Atkinson} except that
the gauge-covariance 
violating term has been removed, since it does not survive 
the four-dimensional momentum integration in the absence of a cut-off. This term 
does not vanish if the angular integral is done first with the radial integral taken 
to infinity afterwards, which is why it must be removed by hand in the cut-off 
approach.

The above equations are finite which can be seen by analysis of the large momentum
limits of the integrands for $Z$ and $M$, which behave asymptotically as
$(k^2)^{-n}$ and $(k^2)^{-r}$ respectively, where $n,r >1$. 
The equation for the wave-function renormalization function
Eq.~(\ref{eq:renwavefunc}) is the same as Eq. (10) in
Ref.~\cite{CP92}, however our treatment of the mass function differs
from theirs.

\subsection{Asymptotic limits of the solutions}
\label{sec:asymp}

In this subsection we solve Eqs.~(\ref{eq:renwavefunc}) and~(\ref{eq:renmass})
analytically for momenta $p^2$ much larger than $M^2(p^2)$ (the asymptotic
region) by linearizing them 
for finite mass, in a manner similar to that used by Atkinson et
al.~\cite{Atkinson} for the chirally symmetric theory (i.e.
$m_0=0$).  It is convenient to {\em temporarily} take the renormalization
point $\mu$ to also be very large, removing this
constraint once our derivation is complete.
The integrals in Eqs.~(\ref{eq:renwavefunc}) and~(\ref{eq:renmass}) are
dominated by contributions around $k^2=p^2$ and $k^2=\mu^2$, so
when $p^2$ and $\mu^2$ are in the asymptotic region
$k^2$ is neccessarily also much greater than $M^2(k^2)$.
Expanding in powers of $M^2(k^2)/k^2$ and keeping at most linear terms
in $M(k^2)$, the kernel functions take their asymptotic forms
$I(p^2,k^2) \rightarrow 0$ and $J(p^2,k^2) \rightarrow J'(p^2,k^2)$,
where $J'(k^2;p^2)$ is defined in appendix~\ref{ap:app1}.
Hence when $p^2$ and $\mu^2$ are both in the asymptotic region, we obtain
\be Z(\mu^2;p^2)&=&1\>-\>\frac{\alpha \xi}{4\pi} \int_{p^2}^{\mu^2}
\frac{dk^2}{k^2} Z(\mu^2;k^2)\quad,
\label{eq:bifwavefunc}  \\    
M(p^2)&=&m_\mu \>+ \> {\alpha \over 4 \pi} 
\int_0^{\infty} d k^2 
\left[ J'(k^2,p^2) - J'(k^2,\mu^2) \right] \;\;.
\label{eq:bifmass}
\ee
These linearized equations are scale invariant and admit power law solutions.

The solution of this asymptotic form of the $Z$ equation is easily
seen to be \cite{DMR}
\be
Z(\mu^2;p^2) \>=\>\left(\frac{p^2}{\mu^2}\right)^{\nu}
\quad\quad \mbox{with} \quad\quad
\nu=\frac{\alpha \xi}{4\pi}\quad,
\label{eq:zsoln_mu}
\ee
which differs from Ref.~\cite{Atkinson} due to the absence of the gauge
covariance violating term.  We thus obtain for $p^2$ and $p'^2$ both
large that
\be
\frac{Z(\mu^2;p^2)}{Z(\mu^2;p'^2)}
 \>=\>\left(\frac{p^2}{p'^2}\right)^{\nu}
\quad\quad \mbox{with} \quad\quad
\nu=\frac{\alpha \xi}{4\pi}\quad,
\label{eq:zsoln_ratio}
\ee
which is valid for an {\em arbitrary} renormalization point $\mu$
since the ratio is by definition renormalization-point independent.
It then follows that for large $p^2$ and any arbitrary $\mu^2$ that
\be
Z(\mu^2;p^2) \>=\>C_\mu \left(\frac{p^2}{\mu^2}\right)^{\nu}
\quad\quad \mbox{with} \quad\quad
\nu=\frac{\alpha \xi}{4\pi}\quad,
\label{eq:zsoln}
\ee
for some appropriate $C_\mu$ such that $C_\mu\to 1$ as $\mu^2$ enters
the asymptotic region.

We shall see that power law solutions of the $M$ equation occur in two
regimes, depending on the value of $\alpha$.  Those with a real exponent
can be identified with the subcritical regime, where $\alpha$ is less
than its critical value $\alpha_c$ which marks the onset of
dynamical chiral symmetry breaking.  Those with complex
exponent correspond to oscillating solutions which correspond to the
supercritical regime where $\alpha>\alpha_c$.
We examine these two cases separately in what follows.

\subsubsection{Subcritical asymptotic solution}

Where, as before, $\mu^2$ and $p^2$ are understood to be chosen  
much greater than all other mass scales, we
try a solution of Eq.~(\ref{eq:bifmass}) of the form
\be
M(p^2)  \>=\>  m_\mu \left(\frac{p^2}{\mu^2}\right)^{-s},\qquad s \in R
\; .
\label{eq:massasym_mu}
\ee
Solving for $s$, we find
\be
&&\frac{3\nu}{2\xi}\Bigg[
                2\pi\cot{s\pi}-\pi\cot{\nu \pi}+3\pi\cot(\nu-s)\pi
                +\frac{2}{(1-s)}+\frac{1}{(\nu+1)}+\frac{1}{\nu}\nonumber\\
&&\hspace{6cm}-\frac{3}{(\nu-s)}-\frac{1}{(\nu-s+1)}\Bigg]
                +\frac{s-1}{\nu-s+1}
                = 0 \quad. 
\label{eq:sdef}
\ee
This equation is the same as in Ref.~\cite{Atkinson}, but in contrast to
the unrenormalized equation examined in that paper, this power Ansatz is
valid for the renormalized equation that we have here whether or not
the bare mass $m_0$ is zero.  We can again form a ratio, i.e.,
provided $p^2$ and $p'^2$ are much larger
than all other mass scales then we have
\be
\frac{M(p^2)}{M(p'^2)}  \>=\>
\left(\frac{p^2}{p'^2}\right)^{-s},\qquad s \in R
\; .
\label{eq:massasym-ratio}
\ee
Both Eqs.~(\ref{eq:sdef}) and (\ref{eq:massasym-ratio}) are manifestly
renormalization-point independent and so are valid for an arbitrary
choice of renormalization point $\mu$.
It then follows that for large $p^2$ and any arbitrary $\mu^2$ that
\be
M(p^2) \>=\>D_\mu \left(\frac{p^2}{\mu^2}\right)^{-s},\qquad s,D_\mu \in R
\label{eq:massasym}
\ee
for some appropriate $D_\mu$.  We see that $D_\mu/(\mu^2)^{-s}$ is
a renormalization-point independent constant and that we must have
$D_\mu\to m_\mu$ as $\mu^2$ moves into the asymptotic region.

There is in fact more than one real
solution to $s$ in Eq.~(\ref{eq:sdef}) 
(see also Ref.~\cite{Atkinson}), e.g., in Landau gauge there are two.
However, only one of these matches
smoothly onto the perturbative solution, and hence is the relevant solution.  
The other solutions appear as a spurious byproduct of the
linearized approximation in Eq.~(\ref{eq:bifmass}) 
(Solutions which do not match smoothly onto
perturbation theory can arise in Eq.~(\ref{eq:bifmass}) if the
integrals diverge like ${1/ \alpha}$, due to infrared divergences; this
cannot occur in  Eq.~(\ref{eq:renmass}) because of the
regulating mass in the denominator).  Since $s$ is real there are no
oscillations in the mass function and hence there is no chiral
symmetry breaking for these couplings. 
While Atkinson et. al. are unable
to generate a mass in this region, due to their initial condition being
$m_0 = 0$, we are able to do so since we are not confined by this
constraint.

\subsubsection{Supercritical asymptotic solution}

In the case where $s$ is complex, the (real) mass function is a superposition
of powers of $s$ and its complex conjugate. For large $p^2$ and
arbitrary $\mu^2$ we have as the counterpart to
Eq.~(\ref{eq:massasym})
\be 
M(p^2) = \frac{1}{2}D_\mu  \left(\frac{p^2}{\mu^2}\right)^{-s} 
         + \frac{1}{2}D_\mu^*\left(\frac{p^2}{\mu^2}\right)^{-s^{*}},
\qquad s, D_\mu \in C
\label{eq:oscil}
\ee 
where $D_\mu$ is some appropriate complex constant, and 
\be
\frac{1}{2}\,
(D_\mu+D_\mu^{*})\,=\,{\rm {Re}}\,(D_\mu)\to m_{\mu}
\ee
as $\mu^2$ enters the asymptotic region.
Here $s$ and $s^*$ are complex conjugate solutions to
Eq.~(\ref{eq:sdef}) and the combination results in a
mass function that oscillates. 
This oscillatory behaviour is more transparent if
we write Eq.~(\ref{eq:oscil}) as
\begin{equation}
M(p^2) = \left( \frac{p^2}{\mu^2} \right)^{-{\rm {Re}}\,(s)}               \left\{ 
                 {\rm {Re}}\,(D_\mu) \cos \left[ {\rm {Im}}\,(s) \log \left(p^2/\mu^2\right) \right]
               + {\rm {Im}}\,(D_\mu) \sin \left[ {\rm {Im}}\,(s) \log \left(p^2/\mu^2\right) \right]  \right\}
  \label{eq:oscil2}
\end{equation}
which has the form of a phase-shifted cosine function periodic in $\log
(p^2/\mu^2)$, with a period of $2\,\pi/{\rm{Im}}\,(s)$, modulated by an exponentially
decaying envelope of $({p^2}/{\mu^2})^{-{\rm{Re}}\,(s)}$.
The oscillations are an indication that chiral symmetry is broken in quenched QED
employing the CP vertex at sufficiently strong coupling $\alpha > \alpha_c$.

\subsection{Numerical solutions}
\label{sec:numerical}

In the regularization-independent approach, the equations are
constructed so as to remove all infinities from the outset
and so the UV region of the integral is not a crucial 
limiting factor.
Moreover, establishing exact asymptotic forms of the wave-function
renormalization and mass functions in the previous subsection
make it possible to analytically integrate
Eqs.~(\ref{eq:renwavefunc}) and~(\ref{eq:renmass}) from the highest
grid point to infinity.  We denote this highest grid point, where the
asymptotic analytic forms are matched onto the numerical results, as the
``matching point'' $k_m^2$.
Hence we can rewrite Eqs.~(\ref{eq:renwavefunc}) and (\ref{eq:renmass}) as
\be 
Z(\mu^2;p^2)  & \equiv &  1 -\frac{\alpha\xi}{4\,\pi}\,\int_{p^2}^{{\mu}^2}
dk^2  \frac{1}{\left[k^2+M^2(k^2)\right]}\,Z(\mu^2;k^2) \nonumber \\
&+& \frac{\alpha}{4\,\pi}
\,\int_0^{k_m^2} 
                \frac{dk^2}{\left[k^2+M^2(k^2)\right]}
    \left [Z(\mu^2;p^2)\,I(k^2,p^2) \> - \> I(k^2,\mu^2) \right ]
\nonumber \\
&+& \,Z_{\rm{high}}(\mu^2;k_m^2,p^2), 
\label{eq:sdexactz}
\\
M(p^2)  & \equiv & m_\mu
+ \frac{\alpha}{4\pi}\int_0^{k^2_m}\frac{dk^2}{\left[k^2+M^2(k^2)\right]}
\left [ J(k^2,p^2) - J(k^2,\mu^2) 
\right.\nonumber \\
&& \left. + \> M(p^2)\>I(k^2,p^2) - m_\mu \>
I(k^2,\mu^2)\right ]
\,+\,M_{\rm{high}}(\mu^2;k_m^2,p^2) \;,
\label{eq:sdexact}
\ee
where the matching point, $k_m^2$, is chosen sufficiently large that the
asymptotic formulas are valid ($k^2_m \gg M^2(k^2_m)$).
These equations serve as definitions of the analytic forms that we need to
evaluate, i.e.,
$Z_{\rm{high}}(\mu^2;k_m^2,p^2)$ and 
$M_{\rm{high}}(\mu^2;k_m^2,p^2)$
are defined as the contributions arising from integrating from $k_m^2$
to infinity for
Eqs.~(\ref{eq:sdexactz}) and (\ref{eq:sdexact}) respectively.
The analytic forms for $Z_{\rm{high}}$ and $M_{\rm{high}}$
and their derivations are given in appendix~\ref{ap:app2}.
By calculating $Z(\mu^2;p^2)$ and $M(p^2)$ 
in this way 
we can achieve very high accuracy in the UV region. 

Eqs.~(\ref{eq:sdexactz}) and (\ref{eq:sdexact}) have been solved 
numerically in Euclidean space for
$Z(\mu^2;p^2)$ and $M(p^2)$ with a variety of gauges $\xi$,
renormalization points $\mu^2$ and renormalized masses
$m_{\mu}$ for  the couplings $\alpha=0.6$ (subcritical) and
$\alpha=1.5$ (supercritical). Each solution was iterated from an
initial guess  until $Z(\mu^2;p^2)$ and $M(p^2)$ converge; our convergence
criteria is that the change
in successive solutions is less than one part in $10^8$ at each momentum point.
Every iteration the data is refitted
to the asymptotic analytic forms of $Z(\mu^2;p^2)$ and $M(p^2)$.

We have found, in line with the above discussion, that the
regularization-independent method allows us to extend
solutions for momenta up to $10^{65}$ as opposed to solutions obtained
from cut-off regularization 
where numerical round-off error meant that the momenta in our solutions 
could typically not exceed ${\cal{O}}(10^{18})$
~\cite{qed4_hw_etal0},~\cite{qed4_hw_etal1},~\cite{qed4_hw_etal2}.
We also verified that our 
solutions do not change when we vary the location of the point $k_m$
where the matching of our numerical and analytical solutions is carried out.

\begin{figure}[H]
\begin{center} 
\epsfig{figure=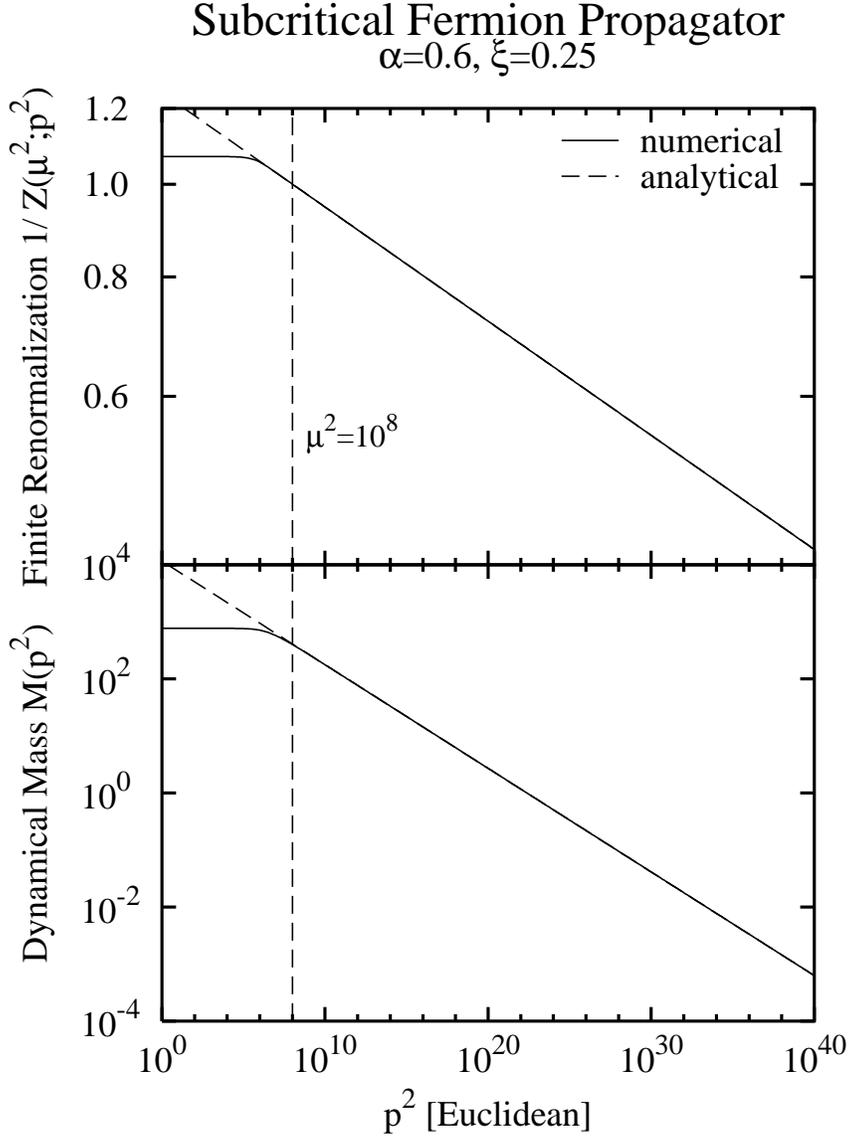,height=15cm} 
\vspace{0.75cm}
\caption{Comparison of the numerical solution for the finite 
         renormalization $1/Z(\mu^2;p^2)$, 
         and the mass function $M(p^2)$ (solid lines) and their predicted
         asymptotic behaviour with matching scales (dashed lines) from
         Eqs.~(\ref{eq:zsoln}) and (\ref{eq:massasym}) 
         for the subcritical coupling $\alpha = 0.6$. 
         The example solution was for a renormalized mass  $m_\mu
	= 400$ (arbitrary units), renormalization point 
	$\mu^2 = 10^8$ and gauge parameter $\xi =0.25$.}
\label{fig:fitsub}
\end{center}
\end{figure}

\begin{figure}[H]
\begin{center}
\epsfig{figure=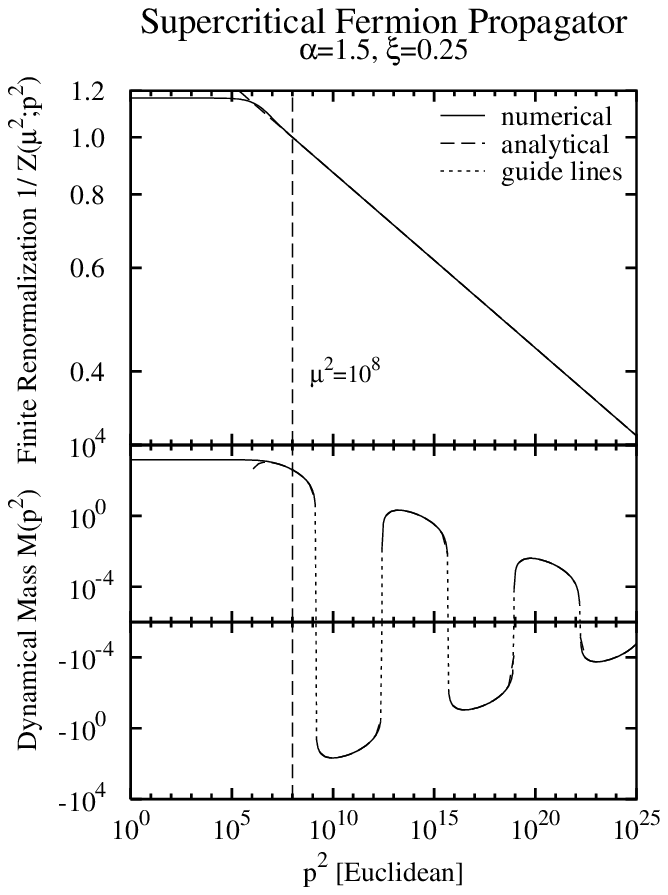,height=15cm} 
\vspace{0.75cm}
\caption{Comparison of the numerical solution for the finite 
         renormalization $1/Z(\mu^2;p^2)$, 
         and the mass function $M(p^2)$ (solid lines) and their predicted
         asymptotic behaviour with matching scales (dashed lines) from
         Eqs.~(\ref{eq:zsoln}) and (\ref{eq:oscil}) 
         for the supercritical coupling $\alpha = 1.5$. 
         The example solution was for a renormalized mass  $m_\mu
	= 400$ (arbitrary units), renormalization point 
	$\mu^2 = 10^8$ and gauge parameter $\xi =0.25$. The `` guide lines'' 
         are added simply to guide the eye between data points on this 
         logarithmic plot }
\label{fig:fitsup}
\end{center}
\end{figure}

In Figs.~\ref{fig:fitsub} and \ref{fig:fitsup}, we show typical solutions,
using regularization independent regularization, 
of the fermion wave-function renormalization and mass functions for
subcritical ($\alpha=0.6$) and supercritical ($\alpha=1.5$) cases
respectively.  Also shown are corresponding solutions based on the theoretical
asymptotic powers with fitted scales.  The log-log
nature of the figures emphasizes the UV behaviour of the propagator.
One can see there is excellent agreement in the region $p^2 \gg M^2(p^2)$.
To emphasize this point we present the theoretical and numerically
calculated powers for these two solutions in Table \ref{tab:table0}.

In summary~: our knowledge of the exact forms of $Z(\mu^2;p^2)$ and $M(p^2)$ 
in the asymptotic region
saves us from the need to extend our numerical solutions far into the
ultra-violet. Therefore Eqs.~(\ref{eq:sdexactz}) and (\ref{eq:sdexact}) 
are solved numerically up to the matching point $(k_m^2)$ in the UV, 
whereafter the analytical solution joins smoothly to the numerical
one. The fitting parameters 
[$C_{\mu}, {\rm{Re}}\,(D_{\mu}), {\rm{Im}}\,(D_{\mu}), {\rm{Re}}\,(s), {\rm{Im}}\,(s), \nu$] 
for the analytic continuation are re-calculated after
each iteration. 

\begin{table}[h]
  \caption{This table compares the powers of the asymptotes of the finite renormalization and mass functions in 
           Figs.~\ref{fig:fitsub} and \ref{fig:fitsup}, as determined analytically and by fitting the numerical
           solutions to a power law form.
           }
\vspace*{0.5cm}
\begin{tabular}{cccccc}
Figure		& coupling	& determination	& $\nu$			& ${\rm{Re}}\,(s)$	& ${\rm{Im}}\,(s)$ \\
\hline\hline
\ref{fig:fitsub}& $\alpha=0.6$	& analytical	& $0.01193662073$	& $-0.181667015$	& \\
		&		& numerical	& $0.01193662073$	& $-0.181666808$	& \\
\hline
\ref{fig:fitsup}& $\alpha=1.5$  & analytical    & $0.02984155182$	& $-0.416012578$	& $0.418128942$ \\ 
                &               & numerical     & $0.02984155183$	& $-0.416012521$	& $0.418129001$ \\
\end{tabular}
\label{tab:table0}
\end{table}

\subsection{Comparison of regularization schemes}
\label{sec:comparison}

We can now compare numerical solutions from the regularization-independent 
approach to those from cut-off regularization with the gauge-covariance 
modification~\cite{qed4_hw_etal0},~\cite{qed4_hw_etal1},~\cite{qed4_hw_etal2}
and those from dimensional regularization~\cite{qed_dim_reg}. 
In cut-off regularization the fermion self-energies are integrated
on a logarithmically spaced grid in $k^2$ momentum up to the highest
(cut-off) momentum $\Lambda^2$. Additionally in dimensional regularization,
an estimate is made of the contribution to the integral from the highest grid
point to infinity. In both the modified cut-off and the dimensional 
regularization  studies subtractive renormalization is performed numerically 
for the regularized (but otherwise divergent) fermion self-energies. For 
the modified UV cut-off regularization approach it was necessary to perform
the calculation at several values of $\Lambda$ and in principle perform a 
$\Lambda \rightarrow \infty $ extrapolation. In practice it was found that it 
was sufficient simply to ensure that $\Lambda$ was chosen large enough. 
For the dimensional regularization approach it was necessary to calculate 
solutions at high accuracy for very many values of $\epsilon$ and then 
carefully extrapolate $\epsilon$ to 0. Numerical limitations made it 
difficult to obtain solutions at very small $\epsilon$, which in turn limit
the achievable accuracy of the $\epsilon \rightarrow 0$ extrapolation. 

\begin{figure}[H]
\begin{center}
\epsfig{figure=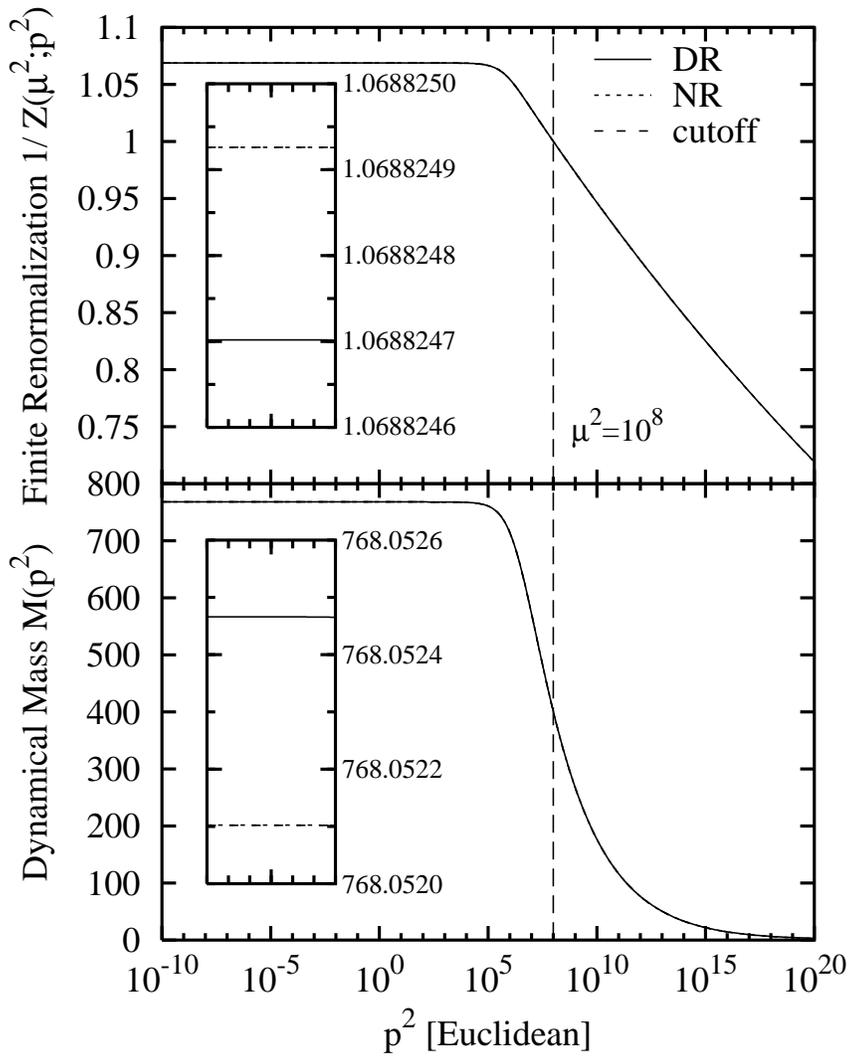,height=15cm} 
\vspace{0.75cm}
\caption{The finite renormalization $1/Z(\mu^2;p^2)$ and dynamical mass
  $M(p^2)$ for the solution of the fermion SDE for subcritical 
  coupling $\alpha=0.6$ and gauge parameter $\xi = 0.25$ found from the
  regularization-independent (NR) method compared with solutions using the
  modified UV cut-off regulator and dimensional regularization. 
  The dimensional regularization solution shown is the result of
  extrapolating various finite $\epsilon$ solutions at scale $10^3$ to
  $\epsilon = 0$ using a fit cubic in $\epsilon$ at each momentum point.
  All solutions have renormalized mass $m_\mu =
  400$ (in arbitrary units) at the renormalization point $\mu^2 = 10^8$.
  The small variation between the three regularization 
  schemes is entirely attributable to limitations achievable in numerical precision.}
\label{fig:numansub}
\end{center}
\end{figure}

\begin{figure}[H]
\begin{center}
\epsfig{figure=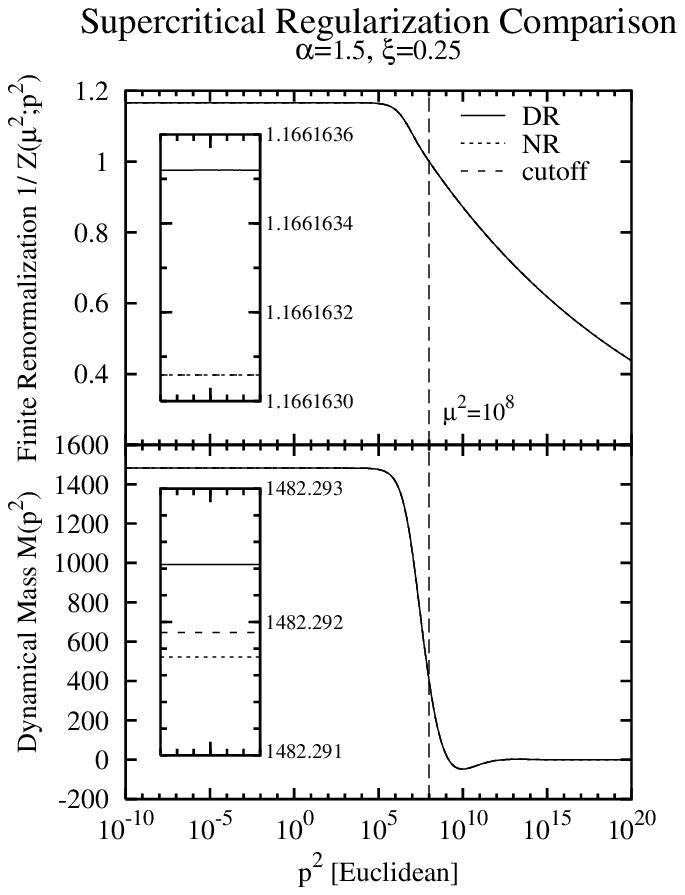,height=15cm} 
\vspace{0.75cm}
\caption{The finite renormalization $1/Z(\mu^2;p^2)$ and dynamical mass
  $M(p^2)$ for the solution of the fermion SDE for supercritical 
  coupling $\alpha=1.5$ and gauge parameter $\xi = 0.25$ found from the
  regularization-independent (NR) method compared with solutions using the
  modified UV cut-off regulator and dimensional regularization. 
  The dimensional regularization solution shown is the result of
  extrapolating various finite $\epsilon$ solutions at scale $10^3$ to
  $\epsilon = 0$ using a fit cubic in $\epsilon$ at each momentum point.
  All solutions have renormalized mass $m_\mu =
  400$ (in arbitrary units) at the renormalization point $\mu^2 = 10^8$.
  The small variation between the three regularization 
  schemes is entirely attributable to limitations achievable in numerical precision.}
\label{fig:numansup}
\end{center}
\end{figure}

In Figs.~\ref{fig:numansub} and \ref{fig:numansup}, we compare these three
regularization methods for subcritical ($\alpha=0.6$) and supercritical ($\alpha=1.5$)
cases respectively, with the standard parameter choice of
$\xi=0.25,\,\mu^2=10^8,\,m_{\mu}=400$.
One can see the agreement between the 
regulator-independent (NR) and the modified cut-off and dimensionally regularized
solutions is excellent. They are indistinguishable on the main
figures: the inserts in Figs.~\ref{fig:numansub} and \ref{fig:numansup}
have the same $p^2$ scale and
reveal the remarkable agreement between them in the infrared region.
In Tables~\ref{tab:table1} and \ref{tab:table2} we quantify the relative 
differences achieved between the three regularization schemes at a variety of 
momentum values. The difference between the results is entirely attributable 
to the limitations achievable in numerical precision.
 We can now conclude with some confidence that 
the three regularization schemes give identical results for the renormalized 
solutions.  



\begin{table}[h]
  \caption{This table shows the relative differences achieved for the finite renormalization
            $1/Z(\mu^2;p^2)$ between the 
           regularization-independent approach and the modified UV cut-off and 
           dimensional regularization (for cubic and quartic fits) approaches.
	   The differences are averages over all momentum points solved in each order of
           magnitude shown. The solution parameters are the same as those in
           Figs.~\ref{fig:numansub} and \ref{fig:numansup}. } 
\vspace*{0.5cm}
 \begin{tabular}{cllllll}
                & reg. indp. vs.          & $p^2 = 10^{-6}$    & $p^2 = 10^{-2}$
                                          & $p^2 = 10^{2} $    & $p^2 = 10^{6} $
                                          & $p^2 = 10^{10}$   \\
  \hline\hline
  $\alpha=0.6$    & mod. cut-off          & $   1.44\times10^{-10}$ & $   1.45\times10^{-10}$
                                          & $   1.94\times10^{-10}$ & $   5.35\times10^{-8} $
                                          & $   8.22\times10^{-8} $ \\
  \hline
                  & dim. reg. (cubic)     & $   2.10\times10^{-7 }$ & $   2.10\times10^{-7 }$
                                          & $   2.09\times10^{-7 }$ & $   9.05\times10^{-8 }$
                                          & $   1.63\times10^{-6 }$ \\
  \hline
                  & dim. reg. (quartic)   & $   9.65\times10^{-8 }$ & $   9.64\times10^{-8 }$
                                          & $   9.63\times10^{-8 }$ & $   6.11\times10^{-8 }$
                                          & $   1.13\times10^{-7 }$ \\
  \hline\hline
  $\alpha=1.5$    & mod. cut-off          & $   1.32\times10^{-10}$ & $   1.73\times10^{-10}$
                                          & $   4.82\times10^{-9 }$ & $   4.97\times10^{-8 }$
                                          & $   2.05\times10^{-7 }$ \\
  \hline
                  & dim. reg. (cubic)     & $   3.95\times10^{-7 }$ & $   3.95\times10^{-7 }$
                                          & $   2.13\times10^{-7 }$ & $   2.23\times10^{-7 }$
                                          & $   3.64\times10^{-6 }$ \\
  \hline
                  & dim. reg. (quartic)   & $   2.04\times10^{-6 }$ & $   2.04\times10^{-6 }$
                                          & $   5.73\times10^{-7 }$ & $   1.27\times10^{-7 }$
                                          & $   6.92\times10^{-7 }$ \\
  \end{tabular}
  \label{tab:table1}
\end{table}

\begin{table}[h]
  \caption{This table shows the relative differences achieved for the mass function
            $M(p^2)$ between the 
           regularization-independent approach and the modified UV cut-off and 
           dimensional regularization approaches. The parameters are the same 
           as those in Figs.~\ref{fig:numansub} and \ref{fig:numansup}.}
\vspace*{0.5cm}
  \begin{tabular}{cllllll}
                  & reg. indp. vs.        & $p^2 = 10^{-6}$    & $p^2 = 10^{-2}$
                                          & $p^2 = 10^{2} $    & $p^2 = 10^{6} $
                                          & $p^2 = 10^{10}$   \\
  \hline\hline
  $\alpha=0.6$  & mod. cut-off            & $   2.63\times10^{-10}$ & $   2.63\times10^{-10}$
                                          & $   3.19\times10^{-10}$ & $   3.51\times10^{-7 }$
                                          & $   1.58\times10^{-6 }$ \\
  \hline
                & dim. reg. (cubic)       & $   4.73\times10^{-7 }$ & $   4.73\times10^{-7 }$
                                          & $   4.73\times10^{-7 }$ & $   4.52\times10^{-7 }$
                                          & $   2.34\times10^{-5 }$ \\
  \hline
                & dim. reg. (quartic)     & $   5.16\times10^{-7 }$ & $   5.16\times10^{-7 }$
                                          & $   5.16\times10^{-7 }$ & $   9.88\times10^{-7 }$
                                          & $   4.18\times10^{-6 }$ \\
  \hline\hline
$\alpha=1.5$	& mod. cut-off            & $   1.24\times10^{-7 }$ & $   1.24\times10^{-7 }$
                                          & $   1.23\times10^{-7 }$ & $   2.57\times10^{-7 }$
                                          & $   6.47\times10^{-5 }$ \\
  \hline
                & dim. reg. (cubic)       & $   4.68\times10^{-7 }$ & $   4.68\times10^{-7 }$
                                          & $   9.29\times10^{-7 }$ & $   1.08\times10^{-6 }$
                                          & $   2.06\times10^{-4 }$ \\
  \hline
                & dim. reg. (quartic)     & $   1.64\times10^{-6 }$ & $   1.64\times10^{-6 }$
                                          & $   5.22\times10^{-7 }$ & $   4.26\times10^{-7 }$
                                          & $   3.74\times10^{-5 }$ \\
  \end{tabular}
  \label{tab:table2}
\end{table}

\section{CONCLUSIONS AND OUTLOOK}
\label{sec:conclusions}

In this paper we have for the first time solved the Schwinger-Dyson 
equations for the fermion propagator in quenched QED$_4$ using the 
regularization-independent approach recently proposed in Ref.~\cite{KSSW}.
This has been done for the particular choice 
of the Curtis-Pennington transverse photon-fermion 
vertex, since this facilitates comparison with previous results   
which used the (gauge-covariance) modified UV cut-off and the 
dimensional regularization schemes. We have carried out precise 
calculations in these three approaches and have achieved excellent
 numerical agreement between them. This clearly demonstrates that we 
are able to achieve high-precision non-perturbative calculations of the 
renormalized fermion propagator that are free from any spurious errors 
which might arise from the regularization procedure itself.

We have derived and used the asymptotic analytic form of the solutions 
to obtain high accuracy even at extremely large momentum scales 
$({\cal{O}}(10^{65}))$. The reason that this is possible now is because 
in the regularization-independent approach all momentum integrations 
are finite by construction, 
[see Eq.~(\ref{eq:submass})]. No bare 
mass or renormalization constants appear in this formulation, since they 
have been eliminated by combining and subracting renormalized quantities.
Since contact with the bare theory is lost, the onset of dynamical 
chiral symmetry breaking is signalled by the onset of oscillations in the 
UV mass function. This is a well-studied phenomena in quenched QED$_4$. 
We have derived the explicit analytical forms for the oscillations in the 
asymptotic region above critical coupling, including the period and decay 
envelope of these. 

The importance of this regularization-independent approach lies in the 
fact that since all unregularized momentum integrations are finite from the outset, 
we do not have the mixing of small and arbitrarily large momentum scales 
in the intermadiate stages of our numerical calculations. This means that 
 we can achieve high accuracy for solutions in the low and medium momentum 
regime with great numerical economy. This new approach will now permit 
numerically tractable studies of {\it unquenched} QED$_4$. These 
studies are now underway.   
\acknowledgements

This work was supported by the Australian Research Council. 
We thank Andreas Schreiber for numerous helpful discussions.

\appendix
\renewcommand{\theequation}{\Alph{section}\arabic{equation}}
\setcounter{equation}{0}
\section{}
\label{ap:app1}
The expressions mentioned in section \ref{sec:sec3} are given below.\\
The Ball-Chiu vertex~\cite{BallChiu} is
\be
\Gamma^\mu_{BC}(\mu;k,p) &=& \frac{1}{2}
\left(\frac{1}{Z(\mu^2;k^2)}
+\frac{1}{Z(\mu^2;p^2)}\right)
\gamma^\mu                      \nonumber \\[3mm] 
&& \hspace{-1.5cm} 
       +\frac{(k+p)^\mu}{k^2-p^2}
        \left[  \left(\frac{1}{Z(\mu^2;k^2)} - \frac{1}{Z(\mu^2;p^2)}\right)
                \frac{({\not\!k}+ {\not\!p})}{2}
              - \left(\frac{M(k^2)}{Z(\mu^2;k^2)} -\frac{M(p^2)}{Z(\mu^2;p^2)}\right)
        \right]\, .
\label{eq:bcvertex}
\ee
The coefficient function of the transverse vertex (Curtis-Pennington)~\cite{mike2} is
\be 
\tau_6(\mu;k,p)=
         -\frac{1}{2\,d}
                   \left(\frac{1}{Z(\mu^2;k^2)} -\frac{1}{Z(\mu^2;p^2)}\right)\quad, \nonumber
\ee
where
\be
d              = \frac{\left\{(k^2 - p^2)^2 +\left(M^2(k^2)  + M^2(p^2)  \right)^2\right\}}{(k^2+p^2)}\quad.
\label{eq:cpvertex}
\ee 
The constituents of the integrand in Eqs.~(\ref{eq:renwavefunc}) and (\ref{eq:renmass}) are the 
renormalization-point independent kernel functions
\be
I(k^2,p^2)&=&\frac{3}{2\,(k^2-p^2)}\,\Bigg\{
                       M(k^2)\left[M(k^2)-M(p^2)\,\frac{Z(\mu^2;k^2)}{Z(\mu^2;p^2)}\right]  \nonumber\\[4mm]
&& \hspace{-1.5cm} 
              + \frac{1}{2}
		\frac{(k^2+p^2)\,(M^2(k^2)+M^2(p^2))^2}{\left\{(k^2-p^2)^2+(M^2(k^2)+M^2(p^2))^2\right\}}\,
                \left[1-\frac{Z(\mu^2;k^2)}{Z(\mu^2;p^2)}\right]\Bigg\}\,
                \left(\frac{k^4}{p^4}\,\theta(p^2-k^2)+\theta(k^2-p^2)\right) \nonumber\\[4mm]
&& \hspace{-1.5cm}             
             +  \xi\,\frac{Z(\mu^2;k^2)}{Z(\mu^2;p^2)}\,\frac{M(k^2)M(p^2)}{k^2}\,
                \left(\frac{k^4}{p^4}\,\theta(p^2-k^2)\right)\quad,
\label{eq:I} \\[5mm] 
J(k^2,p^2)&=&\frac{3}{2}\,M(k^2)\,\Bigg\{
                      1
                    +\frac{Z(\mu^2;k^2)}{Z(\mu^2;p^2)}
                    +\frac{(k^4-p^4)}{\left\{(k^2-p^2)^2+(M^2(k^2)+M^2(p^2))^2\right\}}
                     \left(1-\frac{Z(\mu^2;k^2)}{Z(\mu^2;p^2)}\right)\Bigg\}\, \nonumber\\[4mm]
&& \hspace{5cm}    \times\,\,    \left(\frac{k^2}{p^2}\,\theta(p^2-k^2)+\theta(k^2-p^2)\right)\nonumber\\[4mm]
&-&         \frac{3}{2}\,p^2\, \frac{Z(\mu^2;k^2)}{Z(\mu^2;p^2)}\frac{M(k^2)-M(p^2))}{k^2-p^2}\,
                                \left(\frac{k^4}{p^4}\,\theta(p^2-k^2)+\theta(k^2-p^2)\right)\nonumber\\[4mm]
&+&        \xi\,\frac{Z(\mu^2;k^2)}{Z(\mu^2;p^2)}M(k^2)\left(\frac{k^2}{p^2}\,\theta(p^2-k^2)\right)\quad.
\label{eq:J}
\ee
Linearizing Eqs.~(\ref{eq:I}) 
and~(\ref{eq:J}) in terms of the mass function yields
\be
I(k^2,p^2)\quad &{{\longrightarrow} \atop {{k^2,p^2 \gg M^2}}}& 
\quad I'(k^2,p^2)  \equiv   0  \quad,  \nonumber \\[3mm]
J(k^2,p^2)\quad &{{\longrightarrow} \atop {{k^2,p^2 \gg M^2}}}& 
\quad J'(k^2,p^2)             \quad,   \nonumber \\[3mm]
\mbox{where}\hspace{1.5cm}&&                \nonumber \\
J'(k^2,p^2)&=&
\xi \,\frac{1}{p^2}\,\,\frac{Z(\mu^2;k^2)}{Z(\mu^2;p^2)}\, M(k^2)\,\theta(p^2-k^2) \> \nonumber \\[3mm]
&+& \frac{3}{2}\, 
\Bigg\{ 
          \frac{2 \, M(k^2)}{(p^2-k^2)}\,\,
\left[ p^2 \frac{Z(\mu^2;k^2)}{Z(\mu^2;p^2)}-k^2 \right]
\left[
               \frac{\theta(p^2-k^2)}{p^2}+\frac{\theta(k^2-p^2)}{k^2}
\right]\nonumber \\[4mm]
&-&\hspace{3mm}
 \frac{Z(\mu^2;k^2)}{Z(\mu^2;p^2) }
\frac{ M(p^2)-M(k^2)}{(p^2-k^2)}\,\,
\left[\frac{k^2}{p^2}\,\theta(p^2-k^2)+\frac{p^2}{k^2}\,\theta(k^2-p^2)\right] 
\Bigg \}\;\;.
\ee
%
%
%

\section{}
\label{ap:app2}
This appendix is devoted to the analytic calculation of the
wave-function renormalization and mass functions in the asymptotic
limit. The quantities that we need to evaluate are
\be
Z_{\rm{high}}(\mu^2;k_m^2,p^2)&\equiv& \frac{1}{Z(\mu^2,k_m^2)}
\frac{\alpha}{4\,\pi}\,\int_{k^2_m}^{\infty} \frac{dk^2}{k^2+M^2(k^2)}\,
           \left[Z(\mu^2;p^2)\,I(k^2,p^2)-I(k^2,\mu^2)\right] \nonumber\\[3mm]
M_{\rm{high}}(\mu^2;k_m^2,p^2)&\equiv& \frac{1}{M(k_m^2)}
\frac{\alpha}{4\,\pi}\,\int_{k^2_m}^{\infty} \frac{dk^2}{k^2+M^2(k^2)}\,
                   \nonumber \\
               && \times \left[         J(k^2,p^2)-       J(k^2,\mu^2)
                      +M(p^2)\,I(k^2,p^2)-m_{\mu}I(k^2,\mu^2)   \right]   
\ee
where $I(k^2,p^2)$ and $ J(k^2,p^2)$ are given in appendix~\ref{ap:app1}.
Provided  $k_m^2$ is sufficiently large that $k_m^2\gg M^2(k_m^2)$
then these quantities become
%
%
\be
\int_{k^2_m}^{\infty} \frac{dk^2}{k^2+M^2(k^2)}\,I(k^2,p^2)
  & {=}\atop{k^2_m \gg M^2(k^2_m)} &      
\int_{k^2_m}^{\infty}\frac{dk^2}{k^2}\,
\Bigg\{ \,\,
     - \, \frac{3}{2}\,\,
        \frac{M(p^2)\,M(k^2)}{(k^2-p^2)}\,\,
        \frac{Z(\mu^2;k^2)}{Z(\mu^2;p^2)} \nonumber\\[3mm]
&& \hspace{2cm}            
     + \, \frac{3}{4}\,M^4(p^2)\,\,
        \frac{(p^2+k^2)}{(k^2-p^2)^3}\,
               \left(1-\frac{Z(\mu^2;k^2)}{Z(\mu^2;p^2)}\right) 
\,\, \Bigg\} \nonumber\\[3mm]    
\int_{k^2_m}^{\infty} \frac{dk^2}{k^2+M^2(k^2)}\,J(k^2,p^2)
  & {=}\atop{k^2_m \gg M^2(k^2_m)} &      
\int_{k^2_m}^{\infty} \frac{dk^2}{k^2}\,\Bigg\{\,\,
    3\,M(k^2)
   +3\,M(k^2)\,\frac{p^2}{(k^2-p^2)}\,\left(1-\frac{Z(\mu^2;k^2)}{Z(\mu^2;p^2)}\right)                     
\nonumber\\
&& \hspace{1.6cm}
     -\frac{3}{2}\,\,p^2\,\,\frac{Z(\mu^2;k^2)}{Z(\mu^2;p^2)}\,\frac{M(k^2)-M(p^2)}{(k^2-p^2)} \Bigg\}     
\ee
We have already shown for $k^2\gg M^2$ that the wave-function
renormalization and the mass function have a power law behaviour
\be
Z(\mu^2;k^2) = C_\mu\left(\frac{k^2}{\mu^2}\right)^{\nu} \qquad 
M(k^2)       = \frac{1}{2}D_\mu \left(\frac{k^2}{\mu^2}\right)^{-s} 
              +\frac{1}{2}D_\mu^*\left(\frac{k^2}{\mu^2}\right)^{-s^*}. 
\ee
%
The results for $Z_{\rm{high}}$ and $M_{\rm{high}}$ for large $k^2_m$ and
arbitrary $p^2$ and $\mu^2$ can then be given in terms of hypergeometric functions
\be 
Z_{\rm{high}}(\mu^2;k_m^2,p^2)&=& 
\frac{3\,\alpha}{16\,\pi}\,\left[Z(\mu^2;p^2)
\,\tilde{I}(p^2)-\tilde{I}(\mu^2)\right]
\nonumber\\
M_{\rm{high}}(\mu^2;k_m^2,p^2) 
&=& \frac{3\alpha}{8\,\pi}\, \left[ \tilde{J}(p^2)-\tilde{J}(\mu^2)
+M(p^2)\,\tilde{I}(p^2)-m_{\mu}\,\tilde{I}(\mu^2)
\right] 
\ee
%
%
%
%
with
\be
\tilde{I}(p^2)=
\Bigg\{  &&
       \,\, M^4(p^2) \hspace{1.3cm} \left[ \frac{p^2}{3\,(k^2_m)^3}\,F(3,3,4,p^2/k^2_m) 
                                +          \frac{1}{2\,(k^2_m)^2}\,F(3,2,3,p^2/k^2_m)\right] \nonumber\\
         & -& \frac{M^4(p^2)}{Z(\mu^2;p^2)}\,\,
              \frac{C_\mu}{(\mu^2)^{\nu}}\,\Bigg[\,\,\,
              \frac{1}{(3-\nu)}\,\,
              \frac{p^2}{(k^2_m)^{3-\nu}}\,\,F(3,3-\nu,4-\nu,p^2/k^2_m)  \nonumber\\ 
&& \hspace{3cm}
           +  \frac{1}{(2-\nu)}\,\frac{1}{(k^2_m)^{2-\nu}}\,F(3,2-\nu,3-\nu,p^2/k^2_m)\,\, 
                 \Bigg] \nonumber\\[2mm]
     & +&\frac{M(p^2)}{Z(\mu^2;p^2)}\,\,
         \frac{C_\mu\,D_\mu}{(\mu^2)^{\nu-s}}\,\,
         \frac{(k^2_m)^{\nu-s-1}}{(\nu-s-1)}\,\, F(1,1-\nu+s,2-\nu+s,p^2/k^2_m) \nonumber\\ [3mm]
     &+ &\frac{M(p^2)}{Z(\mu^2;p^2)}\,\,
         \frac{C_\mu\,D_\mu^{*}}{(\mu^2)^{\nu-s^*}}\,\,
         \frac{(k^2_m)^{\nu-s^{*}-1}}{(\nu-s^{*}-1)}\,\, F(1,1-\nu+s^{*},2-\nu+s^{*},p^2/k^2_m) 
\,\,\,\Bigg\}\,,  
\ee
and
\be
\tilde{J}(p^2)=
\Bigg\{  &&
      \frac{D_\mu}{(\mu^2)^{-s}}\,\,
      \frac{p^2}{(1+s)\,(k^2_m)^{s+1}}\,\,F(1,1+s,2+s,p^2/k^2_m)  \nonumber\\
&+&  \frac{D_\mu^{*}}{(\mu^2)^{-s^*}}\,\,
     \frac{p^2}{(1+s^{*})\,(k^2_m)^{s^{*}+1}}\,\,F(1,1+s^{*},2+s^{*},p^2/k^2_m)  \nonumber\\
&-&  
    \frac{3}{2}\,\,
    \frac{C_\mu\,D_\mu}{(\mu^2)^{\nu-s}\,Z(\mu^2;p^2)}\,\,
    \frac{p^2}{(1+s-\nu)\,(k^2_m)^{s-\nu+1}}\,\,
       F(1,1+s-\nu,2+s-\nu,p^2/k^2_m)   \nonumber\\
&-&
    \frac{3}{2}\,\,
    \frac{C_\mu\,D_\mu^{*}}{(\mu^2)^{\nu-s^*}\,Z(\mu^2;p^2)}\,\,
    \frac{p^2}{(1+s^{*}-\nu)\,(k^2_m)^{s^{*}-\nu+1}}\,\,
       F(1,1+s^{*}-\nu,2+s^{*}-\nu,p^2/k^2_m)   \nonumber\\
\nonumber\\
&-&  
    \frac{C_\mu}{(\mu^2)^{\nu}}\,\frac{M(p^2)}{Z(\mu^2;p^2)}\,\frac{p^2}{(\nu-1)\,(k^2_m)^{\nu-1}}\,
       F(1,1-\nu,2-\nu,p^2/k^2_m)
\Bigg\} \,.
\ee
\normalsize


\end{document}